\documentclass[published]{JHEP3} % 10pt is ignored!

\JHEP{00(2007)000}

\JHEPspecialurl{http://jhep.sissa.it/JOURNAL/JHEP3.tar.gz}

\usepackage{epsfig,multicol,bbm}
\newcommand\fverb{\setbox\fverbbox=\hbox\bgroup\verb}
\newcommand\fverbdo{\egroup\medskip\noindent%
                        \fbox{\unhbox\fverbbox}\ }
\newcommand\fverbit{\egroup\item[\fbox{\unhbox\fverbbox}]}
\newbox\fverbbox

\def\be{\begin{equation} }
\def\ee{\end{equation} }
\def\ba{ \begin{eqnarray} }
\def\ea{ \end{eqnarray} }

\def\exd{{\rm d}}
\def\pref#1{(\ref{#1})}

\def\ssA{{\scriptscriptstyle A}}
\def\ssB{{\scriptscriptstyle B}}

\def\ssE{{\scriptscriptstyle E}}

\def\ssR{{\scriptscriptstyle R}}
\def\ssS{{\scriptscriptstyle S}}
\def\ssT{{\scriptscriptstyle T}}

\def\AB{{\scriptscriptstyle AB}}
\def\AcB{{\scriptscriptstyle A,B}}
\def\InfB{{\scriptscriptstyle \infty B}}

\title{Model-Independent Comparisons of \\Pulsar Timings to Scalar-Tensor Gravity}

\author{M.W.~Horbatsch${}^1$ and C.P.~Burgess${}^{1,2}$\\

$^1$ Dept. of Physics \& Astronomy, McMaster University \\
 \qquad 1280 Main St. W, Hamilton, Ontario, Canada, L8S 4L8.\\
\\
$^2$ Perimeter Institute for Theoretical Physics \\
 \qquad 31 Caroline St. N, Waterloo, Ontario, Canada  N2L 2Y5.\\
}

\received{\today}               %%
\accepted{\today}               %% These are for published papers.

\abstract{Observations of pulsar timing provide strong constraints on scalar-tensor theories of gravity, but these constraints are traditionally quoted as limits on the microscopic parameters (like the Brans-Dicke coupling, for example) that govern the strength of scalar-matter couplings at the particle level in particular models. Here we present fits to timing data for several pulsars directly in terms of the phenomenological couplings (masses, scalar charges, moment of inertia sensitivities and so on) of the stars involved, rather than to the more microscopic parameters of a specific model. For instance, for the double pulsar PSR J0737-3039A/B we find at the 68\% confidence level that the masses are bounded by
$1.28 < m_{\ssA}/m_{\odot} < 1.34$ and $1.19 < m_{\ssB}/m_{\odot} < 1.25$,
while the scalar-charge to mass ratios satisfy $|a_{\ssA}| < 0.21$, $|a_{\ssB}| < 0.21$ and $|a_{\ssB}-a_{\ssA}| < 0.002$.
These constraints are independent of the details of the scalar tensor model involved, and of assumptions about the stellar equations of state. Our fits can be used to constrain a broad class of scalar tensor theories by computing the fit quantities as functions of the microscopic parameters in any particular model. 
For the Brans-Dicke and quasi-Brans-Dicke models, the constraints obtained in this manner 
are consistent with those quoted in the literature.
}


\begin{document}

%\maketitle  IS IGNORED %%%%%%%%%%%

\section{Introduction}

Although General Relativity (GR) has many applications in astrophysics and cosmology, incomplete knowledge about the gravitating bodies involved prevents using most of these as tests of the theory itself for the vast majority of systems outside of the solar system. Binary pulsars provide the rare exception to this rule, due to the great precision with which timing measurements allow the properties of their orbits to be inferred \cite{BPrev,Damour:1991rd,PPK}.
These orbital properties are parameterized by the inferred values of the Keplerian parameters (like semi-major axis, $\frak{a}$, and eccentricity, $e$) that define the characteristics of the Newtonian orbit of the pulsar and its partner, together with its orientation in space.
They are also parameterized by a suite of post-Keplerian (PK) parameters that describe observable slow, secular orbital changes over time, as well as
relativistic time delays in the propagation of radio signals emitted by the
pulsar.

\subsection{The classic constraints}

GR predicts the values for these post-Keplerian parameters in terms of the underlying Keplerian parameters and the masses $m_{\AcB}$ of the two orbiting bodies. To test these predictions, bands are drawn in the $m_{\ssA} - m_{\ssB}$ plane, of the form
\be
 \xi^{\rm th}(m_{\ssA},m_{\ssB}) = \xi^{\rm obs} \pm \Delta \,,
\ee
where $\xi^{\rm th}$ denotes the theoretically predicted value of a PK parameter, $\xi^{\rm obs}$ denotes its observed value and $\Delta$ denotes the observational error. The validity of GR requires all such bands to have non-empty intersection, providing a substantive test provided at least three
PK parameters can be measured. The area of mutual overlap then gives the GR-inferred masses of the stars, to within some tolerance.

Thus far, general relativity has passed these stringent tests, using a number of binary pulsars. For two of these --- PSR B1534+12 \cite{1534,1534spin} and the double binary, PSR J0737-3039A/B \cite{wexkramer,dblpsr,dblpsrspin} --- the tests are particularly redundant since it is possible to infer observational values for five independent PK parameters. It is also possible to infer the spin-orbit
precession frequency \cite{1534spin,dblpsrspin}, which can be considered
as a sixth PK parameter.
Moreover, for the double binary, PSR J0737-3039A/B, both stars are pulsars,
and a measurement of the ratio of the semi-major axes yields another
constraint on the masses. The masses are constrained quite accurately
\be
 m_{\ssA} = 1.3381(7) \, m_{\odot}
 \quad \hbox{ and } \quad
 m_{\ssB} = 1.2489(7) \, m_{\odot}
 \quad \hbox{(GR)}\,.
\ee

The success of GR in describing pulsar orbits also constrains alternative theories of gravity; requiring their predictions to agree with GR to within present errors. Prominent among these alternatives are scalar-tensor theories
\cite{sctensrev,defrev}, for which long-range gravitational forces
are mediated by both the metric,
$g_{\mu\nu}$, and a very light scalar, $\phi$, described by the action
\be
 S = - \int \frac{\exd^4 x}{c} \sqrt{-g} \; \left[ \frac{1}{2 \kappa^2} \, g^{\mu\nu}
 \left( R_{\mu\nu} + \partial_\mu \phi \, \partial_\nu \phi \right) \right]
 + S_m \,.
\ee
Here $\kappa^2 = 8 \pi G/c^{4} = \hbar / (M_p^{2} c^{3})$ denotes the gravitational coupling, while $S_m = S_m[g_{\mu\nu}, \phi, \Psi_i]$ denotes the matter action, and controls how $\phi$ and $g_{\mu\nu}$ couple to the various other `matter' fields, $\Psi_i$, which we take in what follows to have the form
\be \label{Smcoup}
 S_m = S_m[ A^{2}(\phi) \, g_{\mu\nu}, \Psi_i] \,.
\ee
This form of coupling has several motivations. First, it is favored by strong observational constraints \cite{EqPrViolations,TEGP} on violations of the weak equivalence principle, which are evaded by actions of the form of eq.~\pref{Smcoup}. Second, it is also the kind of theory that actually describes the low-energy limit of certain types of extra-dimensional theories \cite{XDscalars}.

The predictions of theories of this type have have been compared in detail
with binary pulsar data
\cite{sttest, Damour:2007uf},
with the coupling function assumed to have the particular
{\em quasi-Brans Dicke} form,
\footnote{Brans Dicke theory corresponds to the specific choice $b_s = 0$. Our notation follows that of \cite{HB}, and differs in minor ways from that used elsewhere in the literature \cite{defrev}, as described in detail in table \ref{tblnot}.}
\be
 A(\phi) = \exp \left[ a_{s} \, \phi  + \frac{b_{s}}{2} \, \phi^{2} \right] \,,
\ee
for which the effective coupling of scalars to matter turns out to have the strength
\be
\label{smcoupl}
 a(\phi) := \frac{\exd \ln A }{ \exd \phi}  = a_s + b_s \, \phi \,.
\ee
This form is motivated by the idea that the field $\phi$ does not vary appreciably in any particular system of interest, and so $a(\phi)$ is approximately constant.

\begin{table}[ht]
\begin{center}
\begin{tabular}{c|c|c}
{\bf Our Notation} & {\bf Notation of \cite{defrev}} & {\bf Meaning} \\
\hline\hline
$a_{s}$         & $\alpha$              & Microscopic scalar-matter coupling constant \\
$b_{s}$         & $\beta$               & Microscopic scalar-matter coupling constant \\
$a(\phi)$       & $\alpha(\phi)$        & Microscopic scalar-matter coupling function \\
$a_{\AcB}$      & $\alpha_{\AcB}$       & Effective scalar-matter coupling of a star \\
$b_{\AcB}$      & $\beta_{\AcB}$        & Effective scalar-matter coupling of a star \\
$k_{\AcB}$      & $\kappa_{\AcB}$       & Sensitivity of a star's moment of inertia to scalar field \\
$Q_{\AcB}$      & $\omega_{\AcB}$       & Scalar charge of a star \\
\end{tabular}
\end{center}
\caption{\label{tblnot} Table of Notation.}
\end{table}

Comparison with solar system and pulsar data is found to constrain the microscopic couplings for this theory, $a_s$ and $b_s$, to be consistent with zero, with $a_s$ strongly constrained from solar-system data and $b_s$ restricted by pulsars. The constraints on these microscopic couplings are usually presented as an exclusion plot in the $b_{s}$-$a_{s}$ plane, and in particular it was found that \cite{sttest, Damour:2007uf}:
\be \label{bsbound}
  b_{s} \gtrsim -4.5 \,,
  \qquad
  |a_{s} + b_{s} \phi_{\infty}^{SS}| < 3.4 \cdot 10^{-3} \,,
\ee
where $\phi_{\infty}^{SS}$ is the value of the scalar field
asymptotically far away from the solar system, which is conventionally
taken to be zero.
\subsection{A more model-independent approach}

There are two related drawbacks to traditional comparisons between scalar-tensor theory and observations. First, because limits like eq.~\pref{bsbound} are quoted directly for the microscopic couplings, $a_s$ and $b_s$, a completely new analysis is required for each new assumed functional form for the coupling function $A(\phi)$. Second, as discussed below (and remarked on by the original authors), these bounds are subject to uncertainties that are hard to quantify, due to limits of our understanding of the nuclear equation of state that applies within the pulsars.

One approach towards robustness that has been taken in the literature is to generalize the form taken for the phenomenological lagrangian describing the two-body interactions of the gravitating objects, restricting attention to theories of gravity whose predictions for the orbital dynamics may be derived from a boost-invariant Lagrangian (at least to the first post-Newtonian (1PN) order). Will and Damour \& Taylor have shown that such theories may be characterized by a set of body-dependent phenomenological parameters \cite{TEGP,Damour:1991rd}, of which there are five --- $m_{\ssA}$, $m_{\ssB}$, $\mathcal{G}$, $\epsilon$, $\xi$ --- in the most general Lagrangian \cite{TEGP,Damour:1991rd}:
\be
  \mathcal{L} = \frac{V^{2}}2 + \frac{\mathcal{G}M}{R}
  + \frac{1}{8c^{2}}(1-3\nu)V^{4} +
  \frac{\mathcal{G} M }{2R c^{2}} \left( (\epsilon + \nu)V^{2} + \nu(\mathbf{N} \cdot \mathbf{V})^{2} -
  \xi \frac{\mathcal{G}M}{R}\right) \,,
\ee
where $\mathbf{R}$ and $\mathbf{V}$ are the relative position and velocity vectors, $\mathbf{N} = \mathbf{R}/R$, and
$\mathcal{G}$, $m_{\ssA}$, $m_{\ssB}$, $\epsilon$, and $\xi$ are phenomenological parameters, and $M=m_{\ssA} + m_{\ssB}$, $\nu = m_{\ssA}m_{\ssB}/M^{2}$. Wex and Kramer have used the observed values of the PK parameters for the double
pulsar to constrain these WDT (Will-Damour-Taylor) parameters \cite{wexkramer}. 

However, a drawback of this approach is that it does not describe radiation effects, and in particular the PK parameter $\dot{P}_{b}$ (which describes the shortening of the orbital period due to emission of gravitational radiation). This is because $\dot{P}_{b}$ cannot be expressed in terms of the WDT parameters. This limitation exists because radiative effects come in at higher order in powers of $V/c$ (at 1.5PN order for dipole emission and 2.5PN order for quadrupole emission), whereas the WDT parameters only describe 1PN effects. Extending the phenomenological parametrization to higher PN orders introduces too many new parameters for the data to usefully constrain, however. Thus, in order to use all PK parameters, including $\dot{P}_{b}$, to obtain interesting constraints on alternative theories of gravity, it is necessary to further restrict the class of theories that one considers.

In the present paper we take a complementary approach to confronting pulsar observations with scalar-tensor models, based on the observation that the scalar-tensor predictions for the PK parameters depend only on a relatively small number of macroscopic quantities that characterize how the two stars couple to the scalar field. There turn out to be seven of these, (defined in detail in section \ref{pkpars}): $m_{\AcB}$, $a_{\AcB}$, $b_{\AcB}$ and $k_\ssA$. These generalize the two masses --- $m_\ssA$ and $m_\ssB$ --- that suffice to make predictions within General Relativity, to include four new quantities --- $a_\AcB$ and $b_\AcB$ --- that characterize the strength with which the scalar field couples to the pulsar and its orbital partner, plus one variable, $k_\ssA$, related to the pulsar's moment of inertia.

The key point is that model-dependent complications, like the detailed form of $A(\phi)$ and knowledge of the nuclear equation of state, enter only into the predictions for the quantities $m_{\AcB}$, $a_{\AcB}$, $b_{\AcB}$ and $k_\ssA$ as functions of the microscopic couplings in a particular scalar-tensor theory. But these complications do {\em not} enter at all into the formulae that express how $m_{\AcB}$ through $k_\ssA$ determine the observed PK parameters.

This makes it useful to phrase the confrontation between theory and experiment in two steps: first use the observations to constrain the quantities $m_\AcB$ through $k_\ssA$ once and for all in a model-independent way; then compute these quantities within specific scalar-tensor models as functions of the underlying model parameters that define the function $A(\phi)$.

It is the goal of this paper to perform the first --- and model-independent --- one of these steps. At first sight this might seem to be impossible to do, since it appears to involve constraining more quantities than the five observable PK parameters\footnote{Since the theoretical prediction for the spin-orbit precession frequency has not yet been calculated in scalar-tensor gravity, we do not include it in the present analysis.}. However, it turns out that the present data nonetheless allow useful constraints to be achieved, for two reasons. First, constraints are possible because the dependence of the PK parameters on two of the parameters, $b_{\AcB}$, is very weak; thus effectively reducing the number of free variables from seven down to five.

Second, one combination of Keplerian parameters --- the ratio of projected semi-major axes of the two orbits, $R \equiv x_{\ssB}/x_{\ssA}$ (see section \ref{apdx} for details of notation) --- can sometimes also be measured. For instance for the double pulsar observations give $R = 1.0714 (11)$. This measurement is useful because the theoretical prediction for this quantity in all Lorentz-invariant theories of gravity is $R_{\rm th} = m_{\ssA}/m_{\ssB} + \mathcal{O}(1/c^{4})$ and so is completely independent of the scalar couplings $a_\AcB$ and $k_\ssA$ \cite{Damour:2007uf} (see also section \ref{apdx}). When this ratio is measurable the number of observational constraints to be satisfied rises from five to six.

In the end we find that interesting model-independent constraints on $m_{\AcB}$ and $a_{\AcB}$ are possible. The great virtue of these bounds is that they directly constrain the stellar parameters on which the PK parameters depend, independent of any assumptions about the function $A(\phi)$ and the nuclear equation of state.

Once the observational constraints on these quantities are known, they can be used to constrain the microscopic parameters for any choice of $A(\phi)$ or equation of state as a separate step. When we do so for the quadratic model $A(\phi) = \exp(a_{s} \phi + b_{s}\phi^{2}/2)$, we find agreement with earlier work, which already rules out a large part of the most interesting region (that of spontaneous scalarization \cite{sttest,Damour:2007uf,spsc}).

It might come as a surprise to the reader that this second step might depend on the details of the neutron equation of state, since it is an important feature of general relativity that predictions for the PK parameters depend only on the two masses, $m_{\AcB}$, of the stars, and on none of their other properties. This happy property is called the `principle of effacement' of internal structure \cite{effacement}, and it is this feature that allows a precise prediction of all PK parameters in GR using only the masses, without need for detailed knowledge about the star's structure.

In scalar-tensor gravity (as in most other alternative theories of gravity)
the principle of effacement does not hold. The prediction for the PK parameters
actually depends in principle on all seven of the quantities which characterize the internal gravitational fields: $m_{\AcB}$, $a_{\AcB}$, $b_{\AcB}$, $k_{\ssA}$. For now, it is important to note that these are all a-priori independent. Once a particular equation of state is specified, then the equations of stellar structure can be solved, and on a given branch\footnote{In scalar-tensor theories there may be multiple branches of stellar configurations \cite{spsc}.} of stellar configurations, the quantities $a$, $b$ and $k$ for each star can be expressed in terms of its mass $m$ and the underlying parameters --- like $a_s$ and $b_s$ --- that define the scalar-tensor model. This is the approach taken in \cite{sttest, Damour:2007uf}. The resulting bounds are then subject to uncertainties in the equation of state, which are hard to quantify.

In the end, the phenomenological analysis to which we are led in this paper is similar in spirit to that carried out by Wex and Kramer \cite{wexkramer}. The difference is that we specialize to a more restricted class of theories of gravity, for which the energy loss due to emission of gravitational radiation has been calculated and takes a relatively simple form. Consequently, unlike Wex and Kramer, we are able to
make use of all observed PK parameters, including $\dot{P}_{b}$, when obtaining constraints.

We thus urge observers to express their results in this more model-independent way, which potentially can then be used by theorists to constrain a great variety of specific models. Its independence of the internal structure of the objects involved also shows that some of their physical properties -- like the masses $m_{\AcB}$ -- can be inferred quite robustly, without making assumptions about which theory of gravity actually applies in Nature.

\section{Formalism}

In this section we collect expressions for the Keplerian and Post-Keplerian parameters in scalar-tensor theories, following the results of \cite{Damour:1991rd,PPK} and references therein. We do so both to establish notation and to provide context for the bounds obtained in the next section. Experts and readers in a hurry should feel free to skip this part completely.

\subsection{Keplerian and Post-Keplerian Parameters}
\label{apdx}

First, a reminder of how Keplerian orbits are described, followed by the post-Keplerian parameters that describe slow secular changes to the Keplerian parameters.

\subsubsection*{Orbital description}

The non-relativistic gravitational two-body problem famously predicts bound orbits to be ellipses. More specifically, in the center of mass frame the relative position vector, $\vec{r} = \vec{r}_{\ssA} - \vec{r}_{\ssB}$, sweeps out the trajectory of an ellipse whose shape is described by two parameters: the semi-major axis $\frak{a}$, and the eccentricity $e$. The positions $\vec{r}_{\ssA}$ and $\vec{r}_{\ssB}$ also separately trace out ellipses, with semi-major axes satisfying $\frak{a}_\ssA/\frak{a} = m_\ssB/M$ and $\frak{a}_\ssB/\frak{a} = m_\ssA/M$, where $M = m_\ssA + m_\ssB$ is the system's total mass.

The time taken to traverse this orbit is given by the orbital period $P_{b}$, and is related to the semi-major axis by Newton's modification of Kepler's Third Law:\footnote{$\mathcal{G}$ is the effective gravitational coupling constant between the two bodies.
In scalar-tensor gravity, $\mathcal{G} = G(1+a_{\ssA}a_{\ssB})$.} $P_{b}=2\pi \frak{a}^{3/2}(\mathcal{G}M)^{-1/2}$. For timing measurements a reference time, $T_{0}$, is also needed to specify the time of passage through periastron.

The orientation of the orbital ellipse with respect to a reference triad is specified by three angles. In celestial mechanics these angles are conventionally taken to be the {\it longitude of ascending node}, $\eta$; the {\it orbital inclination}, $i$, of the orbit relative to the plane of the sky; and the {\it argument of periastron}, $\omega$.

Standard techniques allow the positions and momenta of the two bodies to be inferred from the variables $(\frak{a}, e, \eta, i, \omega, T_{0})$ by means of the action-angle formalism \cite{goldstein}.

Deviations from the Newtonian two-body problem cause the shape and orientation of the Keplerian orbits to change with time. In practice this change is slow enough to be regarded as a small secular evolution in each of the Keplerian parameters. Two of these have been accurately measured for several binary pulsars: the decrease in the orbital period due to emission of gravitational radiation, $\dot{P}_{b}$; and the precession of the orbital periastron, $\dot{\omega}$.

\subsubsection*{Pulsar timing}

Pulsars emit electromagnetic signals at regular intervals that are measured
on Earth by radio telescopes. The problem of relating the time of emission $T_{e}$ at the pulsar to the time of arrival $\tau_{a}$ on Earth is usually split into two parts. The first part involves relating $T_{e}$ to the time of arrival $t_{a}$ at the solar-system barycenter, neglecting the time delay due to the solar gravitational field. The second part involves relating $t_{a}$ to $\tau_{a}$, which depends only on the motion of Earth with respect to the solar system barycenter (and is not considered here).

The solution to the first problem is the pulsar timing formula, and is conventionally written
\be
D t_{a} = T_{e} + \Delta_{\ssR} + \Delta_{\ssE} + \Delta_{\ssS} + \mathcal{O}(1/c^{4}) \,.
\ee
These terms have the following origin.
\begin{itemize}
\item The Doppler factor $D$ describes the time dilation caused by the motion of the solar system relative to the binary pulsar.
\item The term $\Delta_{\ssR}$ is called the R\"{o}mer time delay. It is $\mathcal{O}(1/c)$, and is due to the dependence of the light path on the position of the pulsar. The variable that sets the time scale of the R\"{o}mer delay is the projection of the semi-major axis along the line of sight, measured in units of time. It is conventionally denoted $x_{\AcB} \equiv \frak{a}_{\AcB} \sin i / c$, and is called the light crossing time.
\item The term $\Delta_{\ssE}$ is called the Einstein time delay. It is $\mathcal{O}(1/c^{2})$, and is due to the time dilation caused by the motion of the pulsar A and the gravitational field of the companion B. The variable that sets the time scale of the Einstein delay is called $\gamma$, and is given by equation (\ref{pk_gamma}), below.
\item The term $\Delta_{\ssS}$ is called the Shapiro time delay. It is $\mathcal{O}(1/c^{3})$, and is caused by the effect of curved space on the propagation of light. It can be thought of as the first relativistic correction to $\Delta_{\ssR}$. The variable that sets the time scale of the Shapiro delay is called $r$ (the {\it range} of the Shapiro delay), and is given by equation (\ref{pk_r}) below. It is proportional to $G_{\InfB}$, the gravitational coupling between the companion B and a photon, whereas all the other PK parameters depend on $G_{\AB}$, the gravitational coupling between the two orbiting bodies. It is intuitively clear that the importance of the Shapiro effect depends strongly on the inclination angle of the orbit. The variable which parametrizes this dependence is called $s$ (the {\it shape} of the Shapiro delay), and is given by equation (\ref{pk_s}).
\end{itemize}

\subsection{PK Parameters in Scalar-Tensor Gravity}
\label{pkpars}

When solving the problem of stellar structure in scalar-tensor gravity, it is
necessary to specify the boundary condition for the scalar field asymptotically far away from the star, $\phi_{\infty}$. All of the properties of the star, such as its mass, therefore depend implicitly on $\phi_{\infty}$. (Because these properties can be multiple-valued functions of $\phi_\infty$, for some purposes it can be useful instead to follow the dependence on the value of the scalar field at the star's centre \cite{HB}.)

Although the asymptotic field is simple to specify for isolated stars, it is a more complicated concept for an orbiting binary system. For isolated stars, the scalar field sourced by the star becomes small far from its position, leading to the generic weak-field large-distance form
\be
\label{extscfield}
 \Phi_{\AcB}(t,\vec{r}) = \Phi_\infty + \frac{GQ_{\AcB}}
 {|\vec{r}-\vec{r}_{\AcB}(t)|c^{2}}
 + \ldots \,,
\ee
where $Q_{\AcB}$ defines the scalar `charges' of the two stars, and $\vec{r}_{\AcB}(t)$ are their trajectories.\footnote{Knowledge of the stellar equation of state allows $Q_\AcB$ to be computed as functions of the corresponding mass, $m_\AcB$, and the asymptotic value of the scalar field.} We here make the choice that the scalar field approaches $\Phi_\infty$ asymptotically far away from both stars of the binary system. It is convenient to choose units of length so that $\Phi_\infty = 0$.

Now consider a binary system of stars A and B, and assume that the separation
between the two stars is large enough to justify the near-Newtonian weak-field limit. In this regime the fields sourced by the two stars can be superposed so the total scalar field is $\Phi = \Phi_{\ssA} + \Phi_{\ssB}$. Consider now the region much closer to one of the two stars, and let $\phi_{\ssA}$ and $\phi_{\ssB}$ denote the scalar fields of each star in this `internal' regime. The boundary conditions for this `internal' field therefore are (approximately)
\be
\label{scmatch}
 (\phi_\ssA)_{\infty}(t) \simeq \Phi_{\ssB}(t,\vec{r}_{\ssA}(t)) \,, \qquad
 (\phi_\ssB)_{\infty}(t) \simeq \Phi_{\ssA}(t,\vec{r}_{\ssB}(t)) \,.
\ee
These expressions show that $(\phi_\AcB)_\infty$ are small whenever
$Gm/r c^{2} \sim (v/c)^{2} \ll 1$, where $r$ is the separation between the
two stars, and $v$ is their orbital velocity.

Now, the couplings that are relevant for computing post-Keplerian quantities govern how the mass of star A depends on the boundary condition $(\phi_{\ssA})_{\infty}$. For instance, expanding $m_\ssA$ in a power series about $(\phi_{\ssA})_{\infty}=0$, defines the coefficients $a_\ssA$ and $b_\ssA$:
\be
\label{mseries}
 m_{\ssA}[(\phi_{\ssA})_{\infty}] = m_{\ssA} \left[
 1 + a_{\ssA} (\phi_{\ssA})_{\infty} + \frac{1}{2}
 \left( b_{\ssA} + (a_{\ssA})^{2} \right)
 (\phi_{\ssA})_{\infty}^{2} + \ldots \right] \,,
\ee
and an expansion of $m_{\ssB}$ about $(\phi_{\ssB})_{\infty}=0$ similarly defines $a_\ssB$ and $b_\ssB$. We may define the scalar coupling functions, $a_\AcB$ and $b_\AcB$, by
\begin{eqnarray}
 a_{\ssA}[(\phi_{\ssA})_{\infty}] &\equiv& \frac{\partial \log
 m_{\ssA}[(\phi_{\ssA})_{\infty}]}
 {\partial (\phi_{\ssA})_{\infty}}
 = a_{\ssA} + b_{\ssA} (\phi_{\ssA})_{\infty} + \ldots \,,
 \\
 b_{\ssA}[(\phi_{\ssA})_{\infty}] &\equiv& \frac{\partial^{2}
 \log m_{\ssA}[(\phi_{\ssA})_{\infty}]}
 {\partial (\phi_{\ssA})_{\infty}^{2}}
 =b_{\ssA} + \ldots \,,
\end{eqnarray}
with similar definitions for $a_{\ssB}[(\phi_{\ssB})_{\infty}]$ and
$b_{\ssB}[(\phi_{\ssB})_{\infty}]$. It is a general property of scalar-tensor systems that couplings defined in this way agree with those defined from $A(\phi)$ using eq.~\pref{smcoupl} in the limit of weakly coupled non-relativistic systems.

Pulsars rotate, and the frequency of pulsation is given by the rotational frequency $\Omega=J/I$, where $J$ is the `spin' angular momentum of the pulsar, and $I$ is its moment of inertia. $I$ can be found by solving the equations of stellar structure, and just like the mass, it depends on the boundary conditions for the scalar field:
\begin{equation}
 I_{\ssA}[(\phi_{\ssA})_{\infty}] = I_{\ssA}\left[
 1 - k_{\ssA}(\phi_{\ssA})_{\infty} + \ldots  \right]\,,
\end{equation}
leading us to define
\begin{equation}
 k_{\ssA}[(\phi_{\ssA})_{\infty}] \equiv - \frac{\partial \log I_{\ssA}[(\phi_{\ssA})_{\infty}]}
 {\partial (\phi_{\ssA})_{\infty}} = k_{\ssA} + \ldots \,.
\end{equation}

The predictions for the PK parameters in Scalar-Tensor gravity turn out to be determined by the Keplerian parameters, together with the coefficients $m_{\AcB}$, $a_{\AcB}$, $b_{\AcB}$ and $k_{\ssA}$
\cite{Damour:2007uf}. Explicitly,
\begin{eqnarray}
\label{pk_omd}
\dot{\omega} &=& \frac{n}{1-e^{2}}\left(\frac{G_{\AB}Mn}{c^{3}}\right)^{2/3}
\left(
\frac{3-a_{\ssA}a_{\ssB}}{1+a_{\ssA}a_{\ssB}}
- \frac{X_{\ssA}b_{\ssB}a_{\ssA}^{2} + X_{\ssB}b_{\ssA}a_{\ssB}^{2}}
{2(1+a_{\ssA}a_{\ssB})^{2}}
\right) \,,
\\
\label{pk_gamma}
\gamma &=& \frac{eX_{\ssB}}{n(1+a_{\ssA}a_{\ssB})}
\left(\frac{G_{\AB}Mn}{c^{3}}\right)^{2/3}
(X_{\ssB}(1+a_{\ssA}a_{\ssB})+1+k_{\ssA}a_{\ssB}) \,,
\\
\label{pk_r}
r &=& G_{\InfB} m_{\ssB} / c^{3} \,,
\\
\label{pk_s}
s &=& \frac{nx_{\ssA}}{X_{\ssB}}
\left( \frac{G_{\AB}Mn}{c^{3}}\right)^{-1/3} \,,
\end{eqnarray}
where $G_{\AB}=G(1+a_{\ssA}a_{\ssB})$ is the total (graviton plus scalar) weak-field coupling between the two bodies, and $G_{\InfB} = G(1+a_{\infty}^{\rm psr}a_{\ssB})$ is the coupling between the companion $B$ and a non-compact body in the vicinity of the binary pulsar. The quantity $a_{\infty}^{\rm psr}$ is given by the value of the scalar-matter coupling $a(\phi)$ (defined in equation (\ref{smcoupl})) asymptotically far away from the binary pulsar. As before $M = m_{\ssA} + m_{\ssB}$ is the total mass, while $n=2\pi/P_{b}$ is the orbital angular frequency and $X_{\AcB} = m_{\AcB}/M$.

The expression for the decay of the orbital period is similarly given by
\be
 \label{pk_pbd}
 \dot{P}_{b} = \dot{P}_{b}^{\rm mon} +
 \dot{P}_{b}^{\rm dip} + \dot{P}_{b}^{\rm quad} +
 \dot{P}_{b}^{\rm kin} + \dot{P}_{b}^{\rm gal} \,,
\ee
where the different contributions are:
\begin{eqnarray}
 \dot{P}_{b}^{\rm mon} &=& - \frac{3 \pi X_{\ssA}X_{\ssB}}{1+a_{\ssA}a_{\ssB}}
 \left( \frac{G_{\AB} M n}{c^{3}}\right)^{5/3}
 \frac{e^{2}(1+e^{2}/4)}{(1-e^{2})^{7/2}} \times
 \nonumber
 \\
 && \times
 \left[ \frac{5}{3}(a_{\ssA} + a_{\ssB}) - \frac{2}{3}(a_{\ssA}X_{\ssA}
 + a_{\ssB}X_{\ssB})
 +\frac{b_{\ssA}a_{\ssB} + b_{\ssB}a_{\ssA}}{1+a_{\ssA}a_{\ssB}}\right]^{2} \,,
 \\
 \label{diprad}
 \dot{P}_{b}^{\rm dip} &=& - \frac{2 \pi X_{\ssA}X_{\ssB}}
 {1+a_{\ssA}a_{\ssB}}
 \left( \frac{G_{\AB} M n}{c^{3}} \right)
 \frac{(1+e^{2}/2)}{(1-e^{2})^{5/2}} (a_{\ssA}-a_{\ssB})^{2} + \mathcal{O}\left( \frac{1}{c^{5}}\right)\,,
 \\
 \dot{P}_{b}^{\rm quad} &=& - \frac{32 \pi X_{\ssA} X_{\ssB}}
 {5(1+a_{\ssA}a_{\ssB})}
 \left( \frac{G_{\AB} M n}{c^{3}}\right)^{5/3}
 \frac{(1+73e^{2}/24 + 37e^{4}/96)}{(1-e^{2})^{7/2}} \times
 \nonumber
 \\
 && \times
 \left(6 + [a_{\ssA}(1-X_{\ssA}) + a_{\ssB}(1-X_{\ssB})]^{2} \right) \,,
\end{eqnarray}
where `mon', `dip', and `quad' denote monopole, dipole, and quadrupole radiation, respectively. In the limit of general relativity --- {\em i.e.} when $a_{\AcB} \to 0$ and $b_{\AcB} \to 0$ --- the monopole and dipole contributions vanish. Note that the monopole term is of order $1/c^{5}$, because the total scalar charge of the binary system is constant in time.

The last two terms in equation (\ref{pk_pbd}) arise due to the relative motion between the binary pulsar and the solar system \cite{Damour:1990wz}.
The kinetic contribution to $\dot{P}_{b}$ is given by
\begin{equation}
 \dot{P}_{b}^{\rm kin} = v_{\ssT}^{2}/cd \,,
\end{equation}
where $\vec{v}$ is the velocity of the binary pulsar relative to the solar system, and $T$ denotes the component transverse to the line of sight, and $d$ is the distance between the binary pulsar and the solar system.

The galactic contribution to $\dot{P}_{b}$ is given by
\begin{equation}
 \dot{P}_{b}^{\rm gal} = a_{\ssR}/c \,,
\end{equation}
where $\vec{a}$ is the acceleration of the binary pulsar relative to the
solar system, and $R$ denotes the component along the line of sight. We have
\cite{Damour:1991rd,PPK}
\begin{equation}
 \vec{a} = \left[ \frac{1+a_{\infty}^{\rm gal}a_{\rm psr}}
 {1+(a_{\infty}^{\rm gal})^{2}} \right] \vec{g}_{\rm psr}  - \vec{g}_{\rm ss}\,,
\end{equation}
where $\vec{g}_{\rm psr}$ is the acceleration of the binary pulsar relative to the galactic centre, as predicted by a Newtonian galactic model, and $\vec{g}_{\rm ss}$ is the corresponding quantity for the solar system. The quantity $a_{\infty}^{\rm gal}$ is the value of the scalar-matter coupling function $a(\phi)$ (defined in equation (\ref{smcoupl})) asymptotically far away from the galaxy, and
\be
 a_{\rm psr} = X_{\ssA}a_{\ssA} + X_{\ssB}a_{\ssB}
\ee
is the charge-to-mass ratio of the binary pulsar system as a whole.

A simple galactic model may be used to relate $a_{\infty}^{\rm gal}$ to
$a_{\infty}^{\rm psr}$, and so in principle the PK parameters depend on $a_{\infty}^{\rm psr}$ (and the galactic model) in addition to the seven quantities $m_{\AcB}$, $a_{\AcB}$, $b_{\AcB}$ and $k_{\ssA}$. However, it turns out in practice that the contributions of $a_{\infty}^{\rm psr}$ are much too small to be measurable, so we henceforth set $a_{\infty}^{\rm psr} = a_{\infty}^{\rm gal} = 0$. This ensures that GR is recovered asymptotically far away from the binary pulsar, and asymptotically far away from the galaxy.

Note that equations (\ref{pk_omd})--(\ref{pk_pbd}) are invariant under
\be \label{invsym}
 a_{\ssA} \to -a_{\ssA} \,, \quad
 a_{\ssB} \to -a_{\ssB} \,, \quad
 a_{\infty}^{\rm psr} \to -a_{\infty}^{\rm psr} \,, \quad
 a_{\infty}^{\rm gal} \to -a_{\infty}^{\rm gal} \quad \hbox{and} \quad
 k_{\ssA} \to -k_{\ssA} \,,
\ee
which corresponds to switching the sign of the scalar field.

\section{Constraints}

\subsection{Statistics}
In this section, we use the method of least squares \cite{prob} to compare the observed
values of the PK parameters to the predictions of scalar-tensor gravity, thereby obtaining
constraints on the phenomenological stellar parameters $\{ m_{\AcB}, a_{\AcB}, b_{\AcB}, k_{\ssA} \}$.
For brevity, we will denote these phenomenological stellar parameters by $\{\Gamma_{i}\}_{i=1}^{7}$.

For a given pulsar of interest, let $\{\xi_{i}\}_{i=1}^{N}$ run over as many of the quantities
$\dot{P}_{b}$, $\dot{\omega}$, $\gamma$, $r$, $s$ and $R$ as are measured.
Assume that the measurement process for $\xi_{i}$ can be described by a normal (Gaussian)
distribution with standard deviation $\Delta_{i}$, and assume that correlations
between the different $\xi_{i}$ can be neglected.

If we know {\it a priori}
that the pulsar timing is correctly described by the phenomenological stellar parameters
$\{\Gamma_{i} \}$ within scalar-tensor gravity,
then the probability of measuring any given set of values of $\{\xi_{i}\}$ is given by
\be \label{eq:prob1}
P(\xi_1 , \ldots , \xi_{N} | \Gamma_{1} , \ldots , \Gamma_{7} ) =
\frac{e^{-\chi^{2}/2}}{(2 \pi)^{N/2}\Delta_{1} \cdots \Delta_{N}} \,,
\ee
where
\begin{equation}\label{eq:chisq}
\chi^{2}
= \sum_{i=1}^{N} \left[ \frac
{\xi_{i} - \xi_{i}^{\rm th}(\Gamma_{1} , \ldots , \Gamma_{7})}
{\Delta_{i}}\right]^{2} \,,
\end{equation}
where $\xi_{i}^{\rm th}$ are the theoretically-predicted values in scalar-tensor gravity.
By means of Bayes' theorem, this probabilistic statement may be turned around:
if we know {\it a priori} that a given set of values of $\{\xi_{i}\}$ have been observed,
then the probability that the pulsar
is described by the phenomenological parameters $\{ \Gamma_{i} \}$
in scalar-tensor gravity is given by
\be \label{eq:prob2}
P(\Gamma_{1} , \ldots , \Gamma_{7} | \xi_1 , \ldots , \xi_{N} ) =
\frac{P( \Gamma_{1} , \ldots , \Gamma_{7})}{P(\xi_1 , \ldots , \xi_{N})}
P( \xi_1 , \ldots , \xi_{N} | \Gamma_{1} , \ldots , \Gamma_{7}) \,.
\ee
Assume that we have no prior information about $\{\xi_{i}\}$ or
$\{ \Gamma_{i} \}$. Then, combining equations
(\ref{eq:prob1}) and (\ref{eq:prob2}) yields
\be \label{eq:prob3}
P( \Gamma_{1} , \ldots , \Gamma_{7} | \xi_{1}, \ldots , \xi_{N} ) \sim e^{-\chi^{2}/2} \,.
\ee
In principle, equation (\ref{eq:prob3}) may be calculated numerically, and integrated over the
seven-dimensional parameter space to find the constraints of interest. For instance, the mean
value and variance of $\Gamma_{i}$ are given by
\be \label{eq:mamean}
\bar{\Gamma}_{i} = \frac{\int d^{7}\Gamma \, \Gamma_{i} e^{-\chi^{2}/2}}{\int d^{7} \Gamma \, e^{-\chi^{2}/2}}
\ee
and
\be \label{eq:mavar}
\sigma_{\Gamma_{i}}^{2} = \frac{\int d^{7}\Gamma \, (\Gamma_{i} - \bar{\Gamma}_{i})^{2} e^{-\chi^{2}/2}}
{\int d^{7}\Gamma \, e^{-\chi^{2}/2}} \,,
\ee
respectively.
In practice, it is very difficult to calculate (\ref{eq:mamean}) and (\ref{eq:mavar}) directly.
Therefore, we resort to approximation methods.
Assume that $\chi^{2}$ has a global minimum at
$\{ \Gamma_{i}^{\star} \}$, and
approximate it near this minimum by a quadratic form:
\be \label{eq:chi2quadform}
\chi^{2}(\Gamma_{1} , \ldots , \Gamma_{7}) =
\chi^{2}_{\rm min} +
\sum_{i,j=1}^{7} (\Gamma_{i} - \Gamma_{i}^{\star})C_{ij}(\Gamma_{j} - \Gamma_{j}^{\star})
%+ \mathcal{O}(\Gamma^{3})
\,.
\ee
Substituting equation (\ref{eq:chi2quadform}) into equations (\ref{eq:prob3}),
(\ref{eq:mamean}), and (\ref{eq:mavar}) shows that
the $\Gamma_{i}$ are normally distributed, with means
\be
\bar{\Gamma}_{i} = \Gamma_{i}^{\star} \,,
\ee
and variances
\be
\sigma_{\Gamma_{i}}^{2} = \frac{1}{C_{ii}} \,.
\ee
The off-diagonal components of $C$ are related to the correlation coefficients between the different
$\{ \Gamma_{i} \}$.

\subsection{Implementation}

We have implemented the statistical analysis described in the previous section for
two pulsars, for which at least five PK parameters have been measured. For the double pulsar PSR J0737-3039, all of the quantities $\dot{P}_{b}$, $\dot{\omega}$, $\gamma$, $r$, $s$, $R$ have been measured, and so $N=6$. For the binary pulsar PSR B1534+12, by contrast, $R$ is not measured and so $N=5$.

The double pulsar PSRJ0737-3039A/B \cite{wexkramer,dblpsr,dblpsrspin} consists of two pulsars, A and B, that are bound in a relativistic orbit described by the Keplerian parameters summarized in table \ref{tblkep}, and the Post-Keplerian parameters summarized in table \ref{tblpostkep}. Pulsar A has a period of $23 {\rm ms}$, and pulsar B has a much slower period of $2.8 {\rm s}$. The Post-Keplerian timing parameters $\gamma$, $r$, $s$ all pertain to pulsar A. The light crossing time $x$ has been measured for both pulsars.
For the binary pulsar PSR B1534+12 \cite{1534}, the Keplerian parameters are summarized in table \ref{tblkep_1534}, and the Post-Keplerian parameters are summarized in table \ref{tblpostkep_1534}.

\begin{table}[ht]
\begin{center}
\begin{tabular}{c|c|c}
{\bf Symbol} & {\bf Meaning} & {\bf Value} \\
\hline\hline
$x_{\ssA}$ & Light crossing time of A& $1.415032 (1) {\rm s}$ \\
%\hline
$x_{\ssB}$ & Light crossing time of B & $1.5161(16) {\rm s}$ \\
%\hline
$e$ & Eccentricity & $0.0877775(9)$ \\
%\hline
$P_{b}$ & Orbital Period & $0.10225156248(5) {\rm days}$  \\
%\hline
$\omega$ & Argument of periastron of A & $73.805(3)^{\circ}$  \\
%\hline
$T_{0}$ & Time at periastron of A & $52870.0120589(6) {\rm MJD}$
\end{tabular}
\end{center}
\caption{\label{tblkep} Keplerian parameters for the Double Pulsar J0737-3039.}
\end{table}

\begin{table}[t]
\begin{center}
\begin{tabular}{c|c|c}
{\bf Symbol} & {\bf Meaning} & {\bf Value} \\
\hline\hline
$\dot{P_{b}}$ & Time derivative of orbital period & $-1.252 (17) 10^{-12}$  \\
\hline
$\dot{\omega}$ & Precession frequency of periastron
& $16.89947(68)^{\circ} {\rm yr}^{-1}$ \\
\hline
$\gamma$ & Relativistic timing parameter & $0.3856(26)$ \\
\hline
$r$ & Range of Shapiro delay & $6.21 (33) \mu {\rm s}$ \\
\hline
$s$ & Shape of Shapiro delay & $0.99974 (-39,+16)$
\end{tabular}
\end{center}
\caption{\label{tblpostkep} Post-Keplerian Parameters for the Double Pulsar J0737-3039. The galactic and kinetic
contributions to $\dot{P}_{b}$ are negligible. }
\end{table}

\begin{table}[hb]
 \begin{center}
\begin{tabular}{c|c|c}
{\bf Symbol} & {\bf Meaning} & {\bf Value} \\
\hline\hline
$x_{\ssA}$ & Light crossing time of A& $3.729464(2) {\rm s}$ \\
\hline
$e$ & Eccentricity & $0.2736775(3)$ \\
\hline
$P_{b}$ & Orbital Period & $0.420737299122(10) {\rm days}$  \\
\hline
$\omega$ & Argument of periastron of A & $274.57679(5)^{\circ}$  \\
\hline
$T_{0}$ & Time at periastron of A & $50260.92493075(4) {\rm MJD}$
\end{tabular}
\end{center}
\caption{\label{tblkep_1534} Keplerian Parameters for PSR B1534+12.}
\end{table}

\begin{table}[ht]
  \begin{center}
\begin{tabular}{c|c|c}
{\bf Symbol} & {\bf Meaning} & {\bf Value} \\
\hline\hline
$\dot{P_{b}}$ & Time derivative of orbital period & $-0.137(3) 10^{-12}$  \\
\hline
$\dot{\omega}$ & Precession frequency of periastron
& $1.755789(9)^{\circ} {\rm yr}^{-1}$ \\
\hline
$\gamma$ & Relativistic timing parameter & $2.070(2)$ \\
\hline
$r$ & Range of Shapiro delay & $6.7(1.0) \mu {\rm s}$ \\
\hline
$s$ & Shape of Shapiro delay & $0.975(7)$
\end{tabular}
\end{center}
\caption{\label{tblpostkep_1534} Post-Keplerian Parameters for PSR B1534+12. In the value
reported for $\dot{P}_{b}$, the galactic and kinetic contributions have been
subtracted.}
\end{table}

For these two pulsars,
we have calculated (\ref{eq:chisq}) numerically on a grid in the seven-dimensional
parameter space, and found that $\chi^{2}$ has a global minimum near
$a_{\ssA} = a_{\ssB} = 0$. This is not surprising, because these pulsars are very well
described by general relativity. We have also found that the minimum value of $\chi^{2}$ is very
close to zero -- $\chi^{2}_{\rm min} \sim 10^{-3}$ for the double pulsar, and
$\chi^{2}_{\rm min} \sim 10^{-1}$ for 1534+12. This is also not surprising, because we have
more parameters than data points.

Since it is hard to visualize the seven-dimensional parameter space, we will define the following
functions in order to present the results of our numerical calculations:
\ba
\chi_{m}^{2}(m_{\ssA},m_{\ssB}) &\equiv&
\min_{a_{\ssA},a_{\ssB},b_{\ssA},b_{\ssB},k_{\ssA}}
\chi^{2}(m_{\ssA},m_{\ssB},a_{\ssA},a_{\ssB},b_{\ssA},b_{\ssB},k_{\ssA}) \,,
\\
\chi_{a}^{2}(a_{\ssA},a_{\ssB}) &\equiv& \min_{m_{\ssA},m_{\ssB},b_{\ssA},b_{\ssB},k_{\ssA}}
\chi^{2}(m_{\ssA},m_{\ssB},a_{\ssA},a_{\ssB},b_{\ssA},b_{\ssB},k_{\ssA}) \,.
\ea
If we use the quadratic form approximation (\ref{eq:chi2quadform}), then
$\chi_{m}^{2}$ and $\chi_{a}^{2}$ are also quadratic forms, whose coefficients can be explicitly calculated in
terms of the components of $C$ \cite{chisqproof}. Moreover, it can be shown that \cite{chisqproof}
\ba
\label{eq:stddevformula1}
\min_{a_{\ssB}} \chi_{a}^{2}(a_{\ssA}^{\star} + \delta,a_{\ssB})
&=& \chi_{\rm min}^{2} + \delta^{2}/\sigma^{2}_{a_{\ssA}} \,,
\\
\label{eq:stddevformula2}
\min_{a_{\ssA}} \chi_{a}^{2}(a_{\ssA},a_{\ssB}^{\star} + \delta)
&=& \chi_{\rm min}^{2}
+ \delta^{2}/\sigma^{2}_{a_{\ssB}} \,,
\ea
and similar relations hold for $\chi_{m}$. Thus, the standard deviations may be found from contour plots of $\chi_{a}^{2}$ and
$\chi_{m}^{2}$.

\begin{figure}[hb]
\begin{center}
\epsfig{file=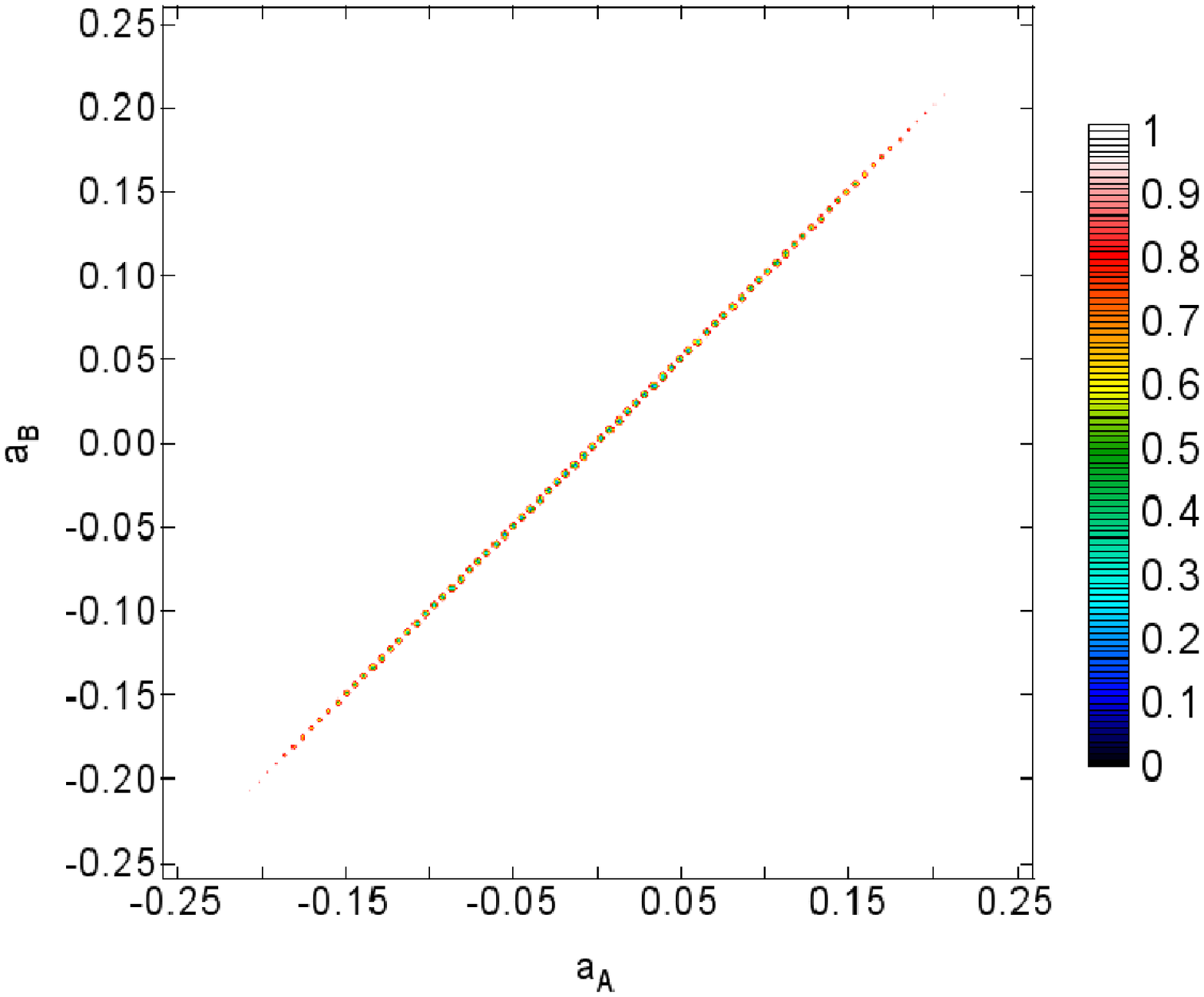,width=0.45\hsize}
\epsfig{file=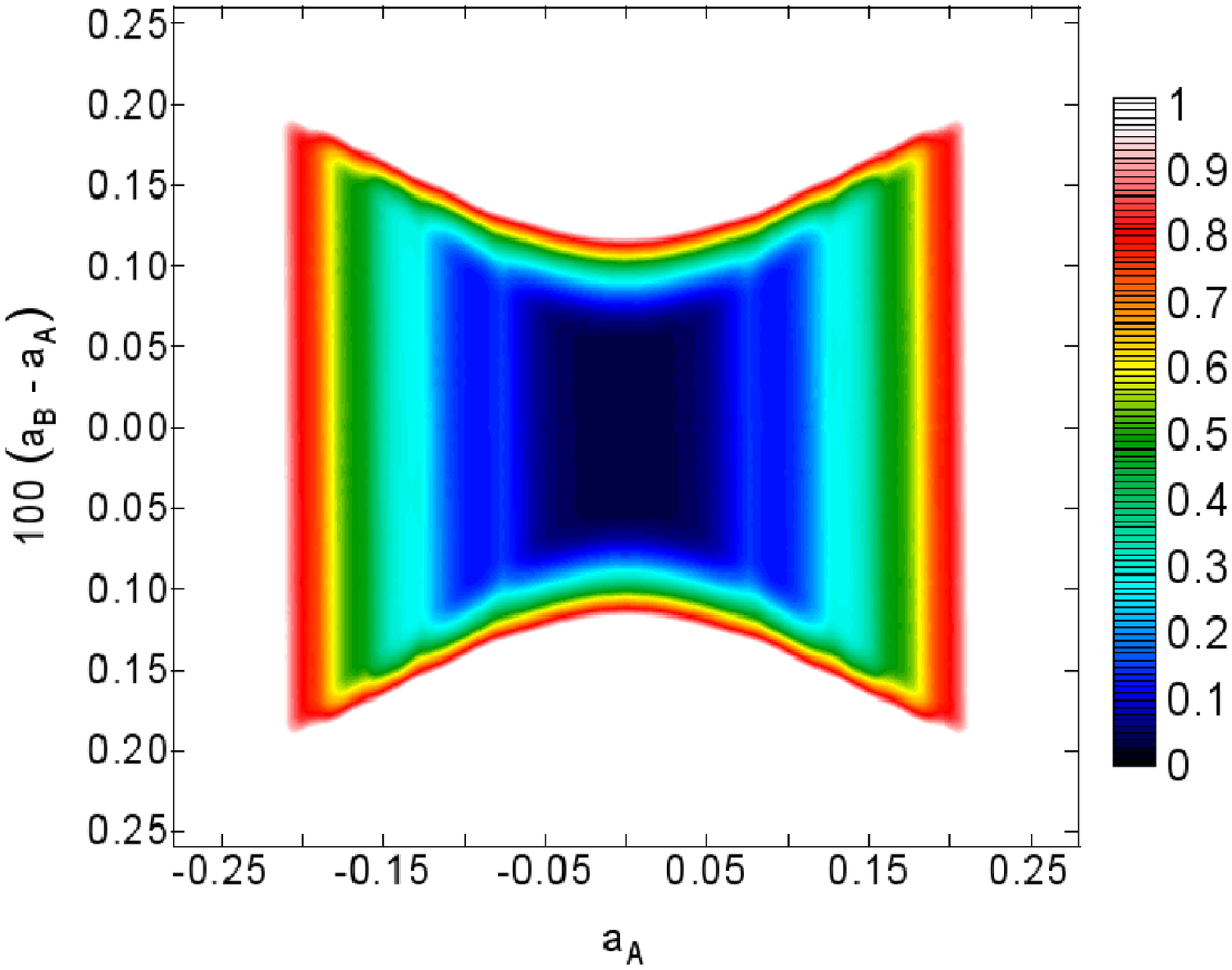,width=0.45\hsize}
\caption{Contour plot of $\chi^{2}$ for the Double Pulsar J0737-3039
as a function of $a_{\ssA}$ and $a_{\ssB}$ (left panel), and a function of $a_\ssB - a_\ssA$ and $a_\ssA$ (right panel). The value of $\chi^{2}$ is minimized over $m_{\ssA},m_{\ssB},b_{\ssA},b_{\ssB},k_{\ssA}$.}
\label{fig:charge.coarse}
\end{center}
\end{figure}

For the double pulsar PSR J0737-3039A/B, contours of $\chi_{a}^{2}$ are plotted in the left panel of figure
\ref{fig:charge.coarse}.
This plot shows that $a_{\ssA}$ is very close to $a_{\ssB}$, with a precision that is more easily seen in the right panel of the same figure. The robustness of this constraint can be simply understood from equation (\ref{diprad}) -- if $|a_{\ssB}-a_{\ssA}|$ gets too large, then this equation predicts too much dipole radiation, inconsistent with observations.
Changing variables from $(a_{\ssA},a_{\ssB})$ to $(a_{\ssA},a_{\ssB}-a_{\ssA})$ in equations
(\ref{eq:stddevformula1})-(\ref{eq:stddevformula2}),
we infer that
\be \label{eq:dblpsr_a_constr}
\sigma_{a_{\ssA}} = 0.21 \,, \qquad
\sigma_{a_{\ssB}-a_{\ssA}} = 0.002 \,,
\ee
while the mean values of both $a_{\ssA}$ and $a_{\ssB} - a_{\ssA}$ are zero.
Also, note that the plots in figure \ref{fig:charge.coarse} are both
symmetric under $a_{\AcB} \to - a_{\AcB}$, as expected from the symmetry (\ref{invsym}).

\begin{figure}[ht]
\begin{center}
\epsfig{file=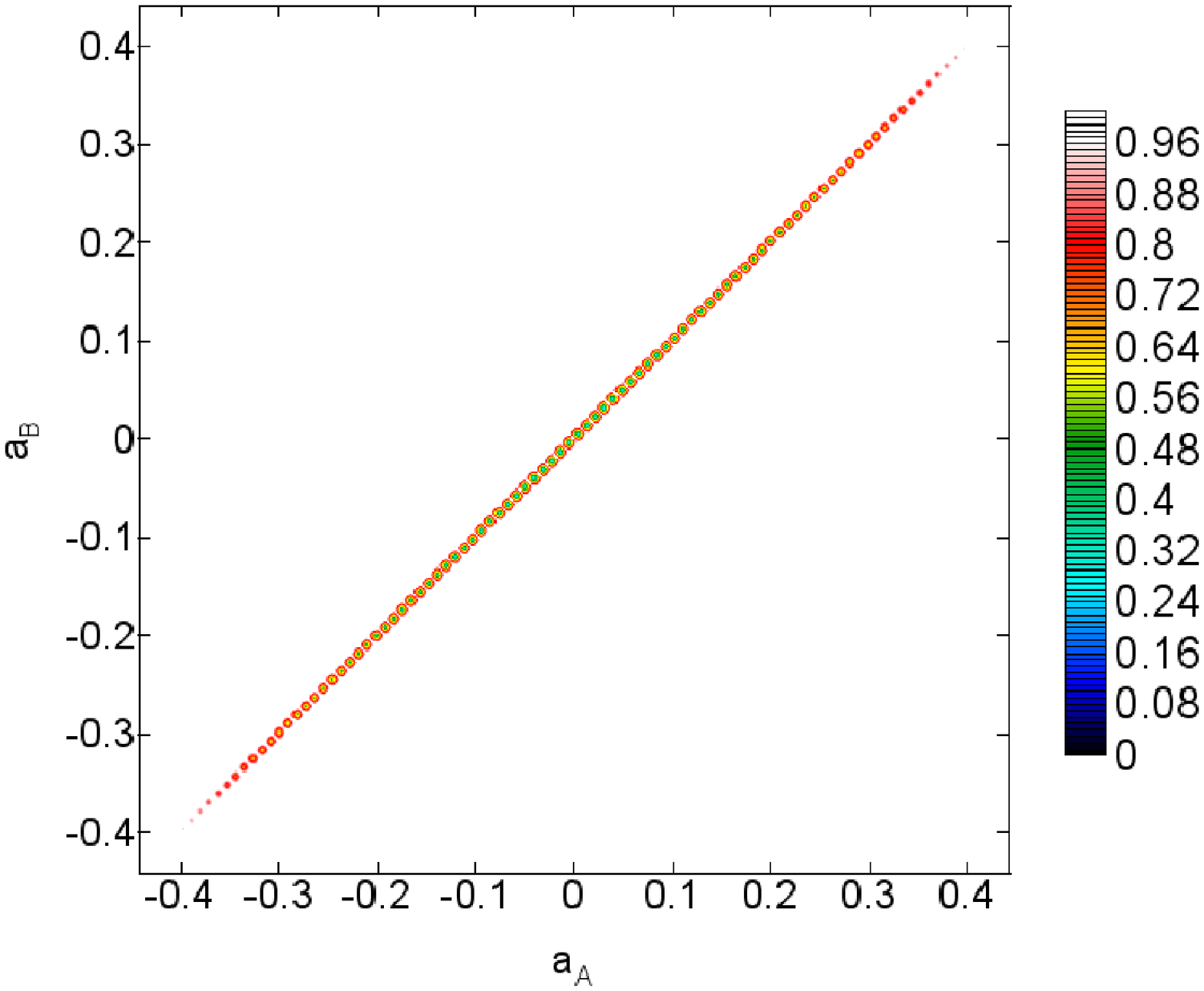,width=0.45\hsize}
\epsfig{file=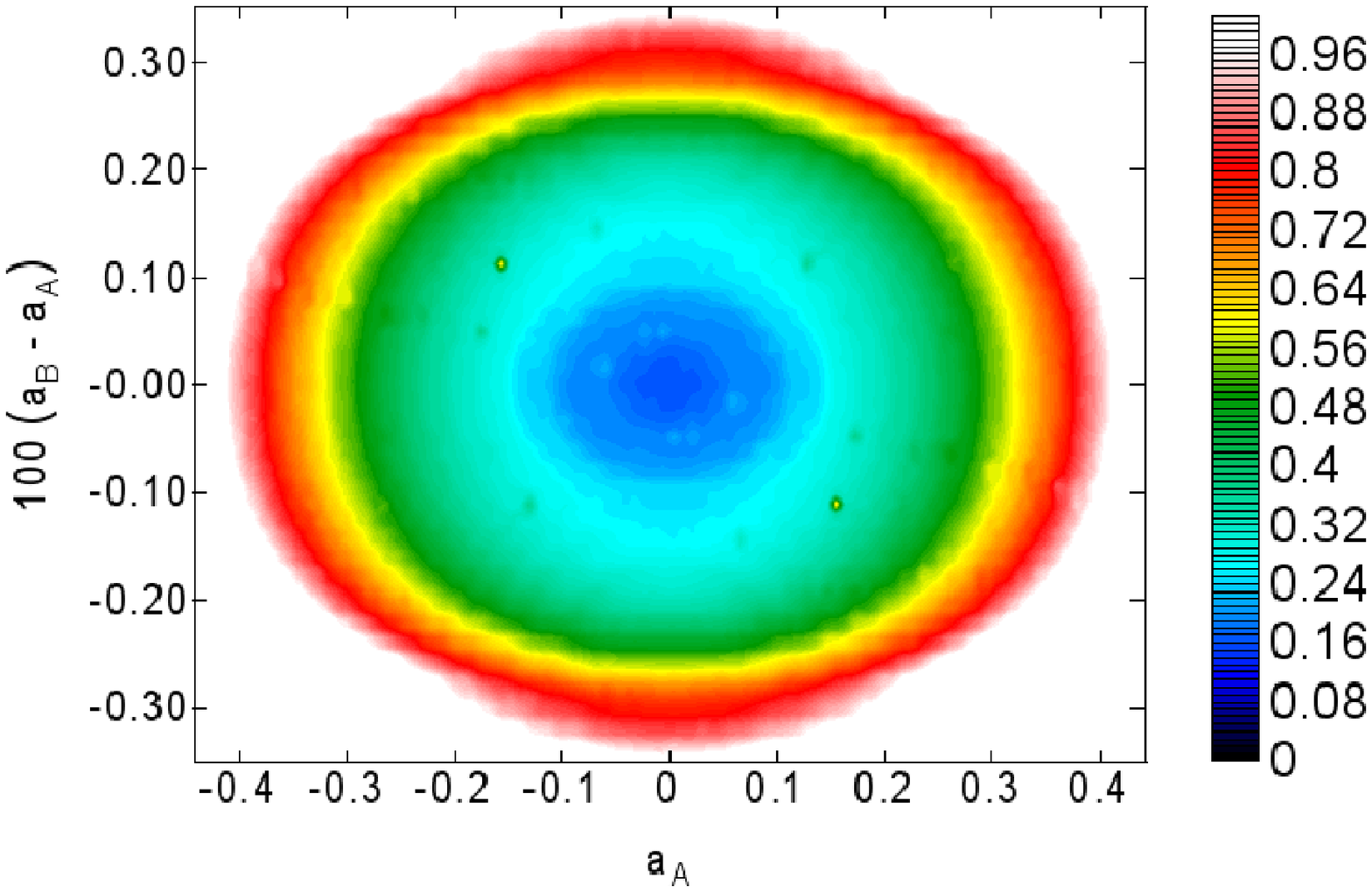,width=0.45\hsize}
\caption{Contour plot of $\chi^{2}$ for PSR B1534+12 in the $a_{\ssA} - a_{\ssB}$ plane (left panel), and with $a_\ssA-a_\ssB$ shown {\em vs} $a_\ssA$ (right panel). The value of $\chi^{2}$ is minimized over $m_{\ssA},m_{\ssB},b_{\ssA},b_{\ssB},k_{\ssA}$.}
\label{fig:charge.coarse_1534}
\end{center}
\end{figure}

This same analysis may be carried out for the pulsar PSR B1534+12. The contour plots of $\chi_{a}^{2}$ are shown in
figure \ref{fig:charge.coarse_1534}. We find that
\be \label{eq:1534_a_constr}
\sigma_{a_{\ssA}} = 0.44 \,, \qquad
\sigma_{a_{\ssB} - a_{\ssA}} = 0.004 \,.
\ee

\begin{figure}[ht]
\begin{center}
\epsfig{file=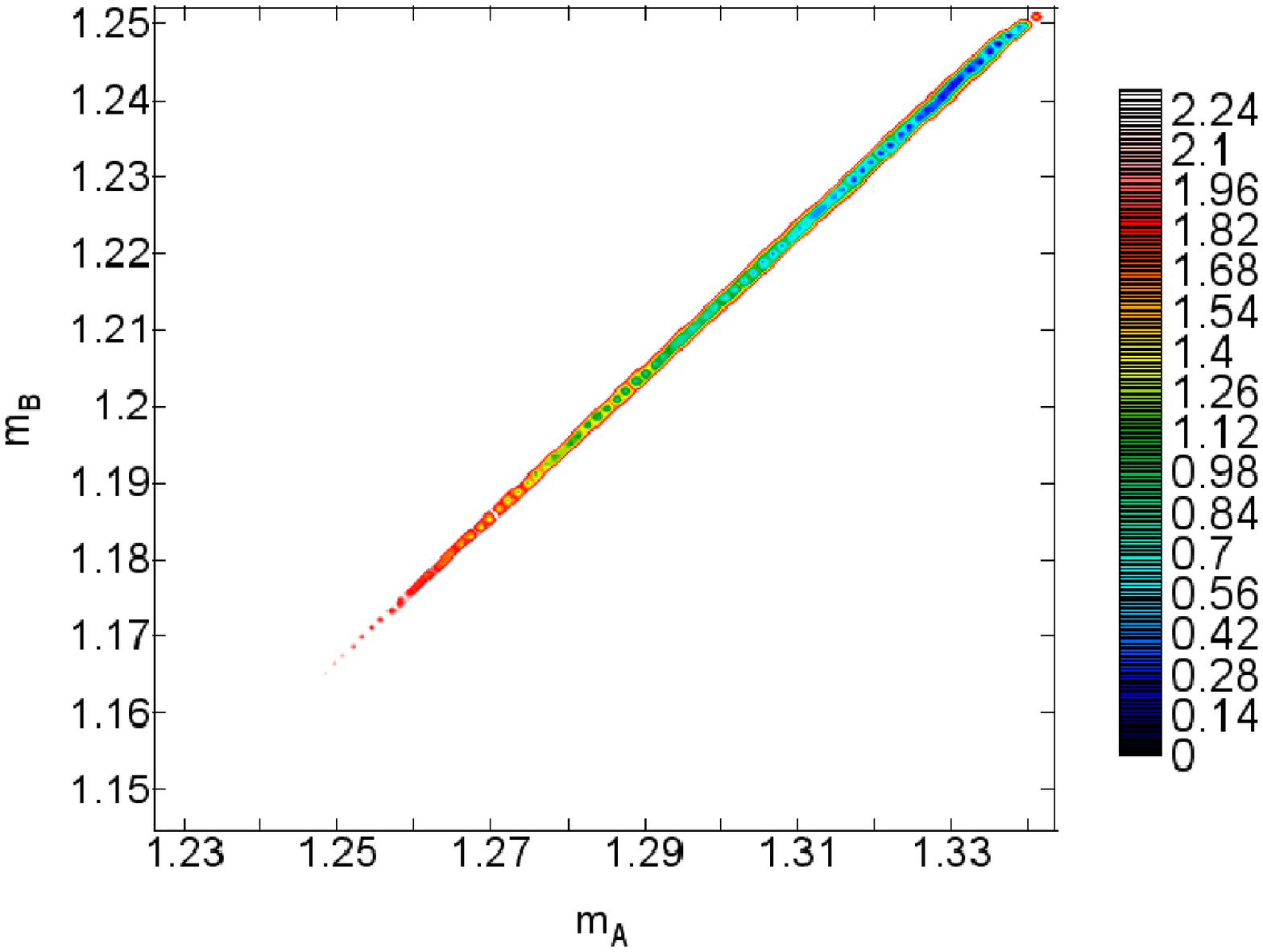,width=0.45\hsize}
\epsfig{file=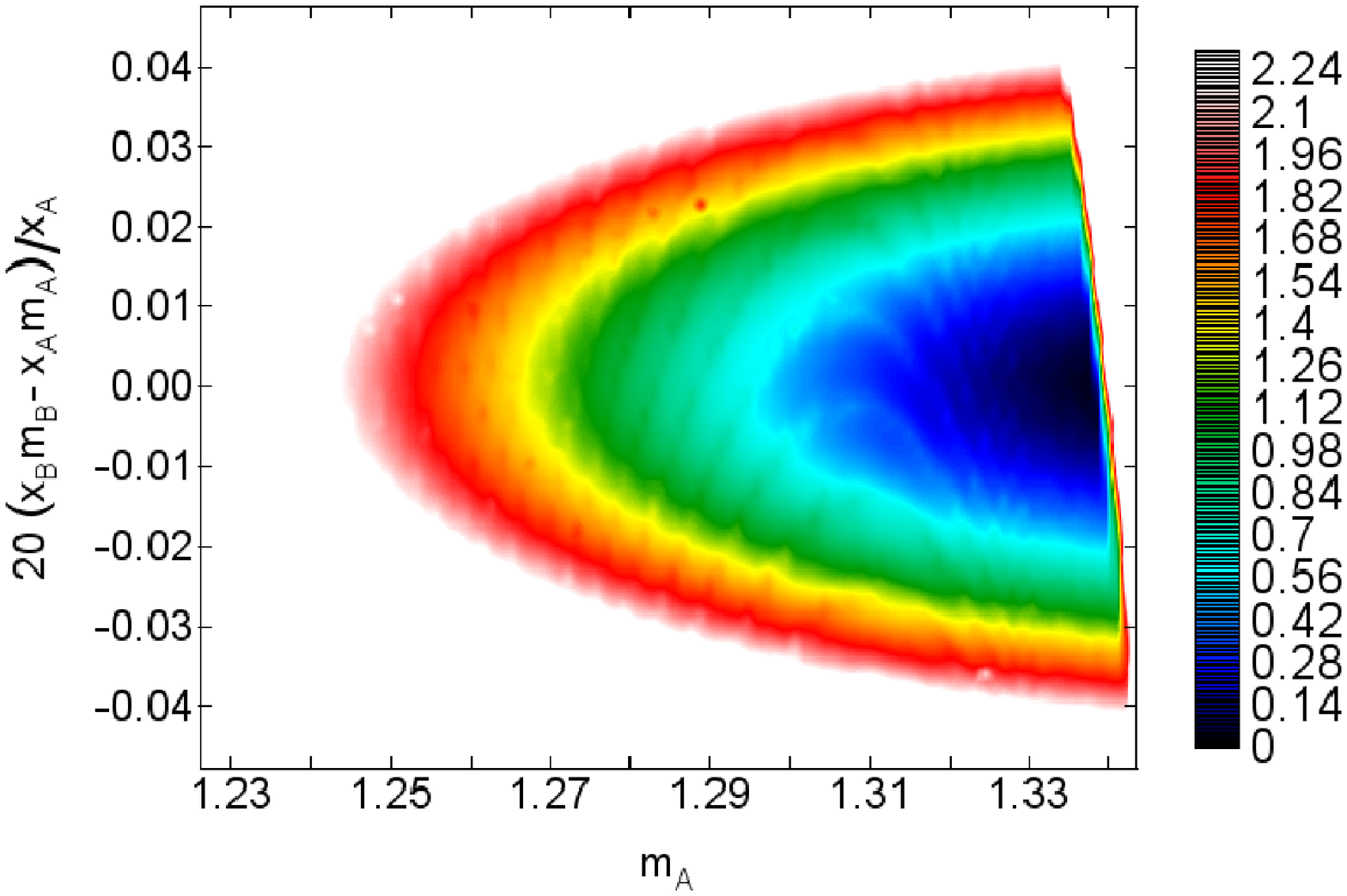,width=0.45\hsize}
\caption{Contour plot of $\chi^{2}$ for the Double Pulsar J0737-3039 in the $m_{\ssA} - m_{\ssB}$ plane (left panel). The right panel plots the same information using a variable that emphasizes the accuracy of the test of the prediction $R = m_\ssA/m_\ssB$. The value of $\chi^{2}$ is minimized over $a_{\ssA},a_{\ssB},b_{\ssA},b_{\ssB},k_{\ssA}$. Units on both axes are in
solar masses.}
\label{fig:mass.coarse}
\end{center}
\end{figure}

For the double pulsar PSR J0737-3039A/B, contours of $\chi_{m}^{2}$ are plotted in the left panel of figure
\ref{fig:mass.coarse}. The masses lie very close to the line $m_{\ssA} = R m_{\ssB}$, in good agreement with the general prediction. The deviation from this line is shown in more detail in the right panel of the same figure, by plotting $x_{\ssB}m_{\ssB} - x_{\ssA}m_{\ssA}$ versus $m_{\ssA}$. The minimum value of $\chi_{m}^{2}$ is close to the mass values inferred in GR --- $m_{\ssA} = 1.3381(7) m_{\odot}$ and $m_{\ssB} = 1.2489(7)m_{\odot}$ --- at the top-right of the line in the left panel of figure \ref{fig:mass.coarse}, and in the center right of the right panel of the same figure.

Note that the behaviour of $\chi_{m}^{2}$ near its minimum value is very asymmetric.
This asymmetry is caused by the contribution of equation (\ref{pk_s}) to $\chi^{2}$. At the
minimum value of $\chi_{m}^{2}$, we have $s \sim 1$. The value of $\chi_{m}^{2}$ increases very rapidly as we
enter the region $s > 1$, whereas $\chi_{m}^{2}$ increases much more slowly as we enter the region $s < 1$. Theoretically, $s$ is the sine of the orbital inclination angle, i.e.
$s = \sin i$ and so $s \leq 1$. Strictly speaking, this
constraint should have been imposed by a prior probability in equation (\ref{eq:prob2}).
However, we see that in practice, for the double
pulsar, this constraint is automatically enforced by the rapid growth of $\chi_{m}^{2}$.

\begin{figure}[hb]
\begin{center}
\epsfig{file=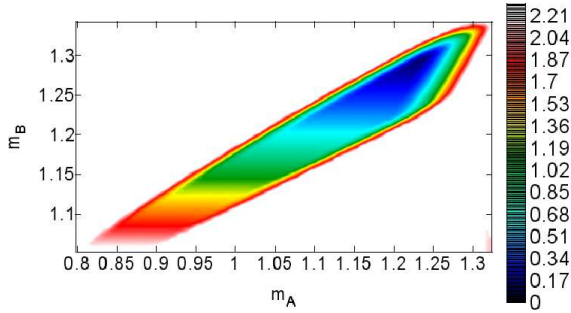,width=0.5\hsize}
\caption{Contour plot of $\chi^{2}$ for PSR B1534+12. The value
of $\chi^{2}$ is minimized over $a_{\ssA},a_{\ssB},b_{\ssA},b_{\ssB},k_{\ssA}$. Units on both axes are in solar masses.}
\label{fig:mass_1534}
\end{center}
\end{figure}

This strong asymmetry also shows that the quadratic form approximation to $\chi^{2}$, eq. (\ref{eq:chi2quadform}),
is not a good one, and that strictly speaking, the use of equations (\ref{eq:stddevformula1})-(\ref{eq:stddevformula2})
is not justified. However, we will still use the $\chi_{m}^{2}=\chi_{\rm min}^{2} + 1$ contour to estimate the allowed
range for the masses at the 68\% confidence level. We find
\be \label{eq:dblpsr_m_constr}
1.28 \leq m_{\ssA}/m_{\odot} \leq 1.34 \,,
\qquad
1.19 \leq m_{\ssB}/m_{\odot} \leq 1.25 \,.
\ee
Now for the pulsar PSR B1534+12, contours of $\chi_{m}^{2}$ are plotted in figure \ref{fig:mass_1534}. The minimum
value is at $m_{\ssA} \sim 1.23 m_{\odot}$, and $m_{\ssB} \sim 1.29 m_{\odot}$.
The masses
are not as highly correlated as in the double pulsar case, because the ratio $R$ has not been measured. Also,
the asymmetry of $\chi_{m}^{2}$ isn't as pronounced as in the double pulsar case, because the PK parameter $s$ has
been measured much less accurately than for the double pulsar. Looking at the contour
$\chi_{m}^{2}=\chi_{\rm min}^{2} + 1$, we find that at the 68\% confidence level,
\be \label{eq:mrange.1534}
0.97 \leq m_{\ssA}/m_{\odot} \leq 1.28 \,,
\qquad
1.15 \leq m_{\ssB}/m_{\odot} \leq 1.31 \,.
\ee
Note that in GR, the masses inferred from the combined measurements of
$\dot{\omega}$, $\gamma$, $r$, and $s$ are $m_{\ssA} = 1.3332(10)m_{\odot}$ and
$m_{\ssB} = 1.3452(10)m_{\odot}$ \cite{1534}, which correspond to the top-right corner of
figure \ref{fig:mass_1534}, and lie outside of the range (\ref{eq:mrange.1534}).
However, including $\dot{P}_{b}$ significantly increases the errors of the GR-inferred masses.
It is often difficult to obtain an accurate measurement of $\dot{P}_{b}$, because this PK parameter
receives galactic contributions.

\clearpage

If we consider the quasi Brans-Dicke model with $A(\phi) = \exp(a_{s} \phi + b_{s} \phi^{2}/2)$,
and use the same relativistic polytrope models for neutron stars as in \cite{spsc} and \cite{HB},
then we find that the double-pulsar constraints
(\ref{eq:dblpsr_a_constr}) and (\ref{eq:dblpsr_m_constr}) together imply that $b_{s} \geq -5.5$ for EOS II,
and $b_{s} \geq -4.6$ for EOS A.
For PSR B1534+12, the constraints (\ref{eq:1534_a_constr}) and (\ref{eq:mrange.1534}) together imply that
$b_{s} \geq -6.2$ for EOS II, and $b_{s} \geq -5.4$ for EOS A. These bounds on $b_{s}$ are similar to those
presented graphically in \cite{Damour:2007uf}.

\section{Conclusions}

In summary, we have shown that the existing data for two pulsars is constraining enough to place model-independent bounds directly on the stellar parameters that control the size of post-Keplerian effects in scalar-tensor models. The virtue of these bounds is that they do not depend on the particular form for the function $A(\phi)$ that defines which model is of interest, or on the details of the stellar equations of state. In particular we find that existing data impose strong and model-independent constraints on the relative size of the two scalar charges and masses.

These bounds can also be used to constrain particular models by computing in these models the masses and couplings as functions of the microscopic parameters. It is at this point that dependence on things like the stellar equation of state enters.

We applied these methods in particular to the double pulsar J0737-3039A/B, and found that $1.28 \leq m_{\ssA}/m_{\odot} \leq 1.34$, $1.19 \leq m_{\ssB}/m_{\odot} \leq 1.25$, $|a_{\AcB}| < 0.21$, and $|a_{\ssB}-a_{\ssA}|< 0.002$, with 68\% confidence, for all choices of scalar-matter coupling function, and for all nuclear equations of state.

A similar analysis of the pulsar PSR B1534+12 yields $0.97 \leq m_{\ssA}/m_{\odot} \leq 1.28$, $1.15 \leq m_{\ssB}/m_{\odot} \leq 1.31$, $|a_{\AcB}| < 0.44$, and $|a_{\ssB}-a_{\ssA}|< 0.004$.

\section*{Acknowledgements}

We wish to thank Marko Horbatsch for discussions and advice on numerical
algorithms and contour plots, Maxim Lyutikov and an anonymous referee for useful comments, and SHARCNET for computational resources. Our research is supported in part by funds from the Natural Sciences and Engineering Research Council (NSERC) of Canada. Research at the Perimeter Institute is supported in part by the Government of Canada through Industry Canada, and by the Province of Ontario through the Ministry of Research and Information (MRI).

\end{document}